\documentclass[12pt,preprint]{aastex}
\usepackage{epsfig}
\usepackage{graphicx}
\usepackage{color}
\usepackage{enumerate}
\usepackage{ulem}
\usepackage{epsfig}

\newcommand{\accrate}{{\rm\,M_\odot~yr^{-1}}}

\def\beq{\begin{equation}}
\def\eeq{\end{equation}}

\def\kms{{\rm\,km\,s^{-1}}}
\def\kpc{{\rm\,kpc}}
\def\msun{{\rm\,M_\odot}}

\def\rsun{{\rm\,R_\odot}}
\def\pc{{\rm\,pc}}

\def\yr{{\rm\,yr}}
\def\Myr{{\rm\,Myr}}

\def\Mbh{M_{\rm BH}}

\shorttitle{INTERACTION OF RECOILING BLACK HOLES WITH STARS} \shortauthors{Li et al.}

\begin{document}

\title{INTERACTION OF RECOILING SUPERMASSIVE BLACK HOLES WITH STARS IN
GALACTIC NUCLEI}

\author{Shuo Li\altaffilmark{1,4}, F. K. Liu\altaffilmark{1}, Peter Berczik\altaffilmark{3,4,5},
  Xian Chen\altaffilmark{2}, and Rainer Spurzem\altaffilmark{3,4,2} }

\altaffiltext{1}{Astronomy Department, Peking University, 100871 Beijing, China;
 fkliu@bac.pku.edu.cn, lis@bac.pku.edu.cn, chenx@bac.pku.edu.cn}
\altaffiltext{2}{Kavli Institute for Astronomy and Astrophysics, Peking University, 100871 Beijing, China}
\altaffiltext{3}{National Astronomical Observatories of China, Chinese Academy of Sciences, 20A Datun Lu, Chaoyang District, 100012, Beijing, China}
\altaffiltext{4}{Astronomisches Rechen-Institut, Zentrum f\"{u}r Astronomie, Universit\"{a}t Heidelberg, M\"{o}nchhofstr. 12-14, D-69120 Heidelberg, Germany}
\altaffiltext{5}{Main Astronomical Observatory, National Academy of Sciences of Ukraine, 27 Akademika Zabolotnoho St., 03680 Kyiv, Ukraine}

\begin{abstract}

Supermassive black hole binaries (SMBHBs) are the products of frequent galaxy mergers. The coalescence of the SMBHBs
 is a distinct source of gravitational wave (GW) radiation. The detections of the strong GW radiation and their possible
electromagnetic counterparts are essential. Numerical relativity suggests that the post-merger supermassive black hole
(SMBH) gets a kick velocity up to $4000\kms$ due to the anisotropic GW radiations. Here we investigate the dynamical
co-evolution and interaction of the recoiling SMBHs and their galactic stellar environments with one million direct
$N$-body simulations including the stellar tidal disruption by the recoiling SMBHs. Our results show that the accretion of
disrupted stars does not significantly affect the SMBH dynamical evolution. We investigate the stellar tidal disruption
rates as a function of the dynamical evolution of oscillating SMBHs in the galactic nuclei. Our simulations show that most
of stellar tidal disruptions are contributed by the unbound stars and occur when the oscillating SMBHs pass through the
galactic center. The averaged disruption rate is $\sim10^{-6} \accrate$, which is about an order of magnitude lower than
that by a stationary SMBH at similar galactic nuclei. Our results also show that a bound star cluster is around the
oscillating SMBH of about $\sim 0.7\%$ the black hole mass. In addition, we discover a massive cloud of unbound stars following the oscillating SMBH. We also investigate the dependence of the results on the SMBH masses
and density slopes of the galactic nuclei.
\end{abstract}

\keywords{galaxies: active --- galaxies: evolution --- galaxies: kinematics and dynamics --- galaxies: nuclei ---
methods: numerical}

\section{INTRODUCTION}
\label{introduction}

Supermassive black hole binaries (SMBHBs) are predicted by the hierarchical galaxy formation model in $\Lambda$ cold
dark matter ($\Lambda$CDM) cosmology \citep{beg80,vol03}. For merging galaxies, their two SMBHs with galactic cores
will firstly approach each other by dynamical friction, and then get close enough to form a bound, compact binary
system. After that, the SMBHB may stall at a so called ``hard binary separation'' for a time even longer than the
Hubble time \citep{beg80}. However, recent investigations suggest that the hardening rates of SMBHBs can be boosted and
they may coalesce within the Hubble time either due to various stellar dynamical processes other than spherical
two-body relaxation \citep{yu02,cha03,mer04,ber06,ses08}, or gas dynamics \citep[][and references
therein]{gou00,cop09}.

Since the evolution of SMBHBs deeply impacts the evolution of host galaxies, it is very important for us to find observational
evidences to constrain evolution models of SMBHBs. A statistic way is the calculation of tidal disruption rates. A dormant
SMBH could be temporarily activated by tidally disrupting a star passing by and accreting the disrupted stellar debris
\citep{hil75,ree88,eva89,lod09}, which may have been observed in several non-active galaxies \citep{kom99,kom02,gez08,gez09}.
\citet{che08} and \citet{che09} calculated the tidal disruption rate in SMBHB systems at different evolutionary stages, and found
that is significantly different from the typical rate in a single SMBH for several orders of magnitude.

The most certain proof for detecting SMBHB individually may come from gravitational wave (GW) radiation observation. Coalescing
SMBHBs are important sources of GW radiation \citep{pet64,beg80}, and can be detected within the next decades by the Laser
Interferometer Space Antenna (LISA) and the Pulsar Timing Array (PTA) program \citep{bere09,ama10}. Because of the poor spatial
resolution of both LISA and PTA for locating GW radiation sources, it is of key importance to detect electromagnetic counterparts
(EMCs) of GW radiation sources. Besides, identifying SMBHBs by their EMCs is also essential to constrain the poorly understood
galaxy-merger history.

Several EMCs have been suggested in the literature to probe SMBHBs, (1) precession of jet orientation and its
acceleration in radio galaxies during the in-spiraling of SMBHBs \citep{beg80,liu07}, (2) optical periodic outbursts in
blazars due to the interaction of SMBHBs and accretion disk \citep{sil88,liu95,liu06,liu02,val08,hai09}, (3) jet
reorientation in X-shaped radio galaxies due to the exchange of angular momentum between SMBHBs and accretion disk
\citep{liu04}, (4) interruption of tidal disruption flares in SMBHBs systems \citep{liu09}. Besides, there are also
some EMCs to probe the coalescence of SMBHBs and its remnant, (1) intermittent activity in double-double radio galaxies
at binary coalescence \citep{liu03}, (2) X-ray, UV, optical, and IR afterglow following binary coalescence
\citep{mil05,shi08,lip08,sch08}, (3) systematically shifted broad emission lines relative to narrow emission lines
\citep{kom08a} and off-center active galactic nuclei (AGNs) \citep{mad04,loe07} because of SMBH GW radiation recoil,
(4)tidal disruption flares and hypercompact stellar systems for recoiling black holes \citep{kom08b,ole09,mer09,sto10}.

The breakthrough on numerical relativity in the past few years reveals the final stage of SMBHB's coalescence
\citep{pre05,cam06,bak06a}. The coalescence remnant SMBH can be recoiled due to the anisotropically GW emission during the
inspiral and final coalescence \citep{per62}. For nonspinning SMBHs, the recoil velocity is $V_{\rm kick}\lesssim 200\kms$
\citep{bak06b,gon07a,her07}, which is just as the same order of magnitude as the stellar dispersion velocity in the galactic center.
However, if both of two SMBHs are rapidly spinning, the recoil velocity can be as large as thousands kilometers per second.
This recoil velocity depends sensitively on the intersection angle between spin vectors of two SMBHs and their linear momenta.
For some extreme cases, with maximally spinning equal-mass SMBHs and antialigned spins in orbital plane, the recoiling
velocity can achieve to $\sim 4000\kms$ \citep{cam07,gon07b}.

A recoiling SMBH with high velocity has significant displacement relative to the galactic center. For most of massive
galaxies with escape velocity $\lesssim 3000\kms$ \citep{mer04a}, the recoiling SMBH has opportunity to escape from its host
galaxy. Those recoiling SMBHs which are still bound to host galaxies will oscillate around the galactic centers and change the core
density profiles of the host galaxies due to dynamical friction \citep{boy04,gua08}. In addition, the recoiling SMBH with a fraction of
its accretion disk can produce sets of emission lines separated by relative velocity \citep{kom08a} or off-center active galactic
nuclei (AGNs) \citep{mad04,loe07}.

In gas poor environment, a recoiling SMBH might be observed as a ``hypercompact stellar system'' (HCSS). That is a kind of star cluster which is bound to the recoiling SMBH, with similar luminosity as a globular cluster or ultracompact dwarf galaxy, and very high velocity dispersion \citep{mer09,ole09}. The other signature is an offset tidal flare due to the recoiling SMBH tidally
disrupting surrounding stars. \citet{kom08b} have used both analysis and $N$-body simulations to calculate the tidal disruption
rates of the recoiling SMBH for bound and unbound stars. They found that the tidal disruption rates, in most cases, are
smaller than in stationary systems. Recently, \citet{ole11} have investigated the bound stellar density profile
and tidal disruption evolution around the recoiling SMBH. They found that the tidal disruption rate will monotonically fall as a
power-law after the recoil. There is also an analytical work focusing on the tidal disruption rate during a short period just after
coalescence \citep{sto10}, which predicts multiple tidal disruption flares in few years or decades after the coalescence.

Most of the above works about tidal disruption of the recoiling SMBH have focused on theoretical analysis and neglected the impact of
background stars. Since the real evolution process in such kind of system is very complex, a detailed study on the recoiling
SMBH co-evolving with the galactic center is essential. Unlike previous work, we evolve the entire system before and after the recoil
with unbound stars included. Our work focuses on investigating the co-evolution between the recoiling SMBH and surrounding stars with
tidal disruption and accretion processes. To it come true, we use a special high-accuracy, parallel direct $N$-body code
accelerated by many-core hardware ($\varphi$\,-{\sc Grape} and $\varphi$\,-GPU) \citep{ber05,har07,spu09,jus11}, including a simplified tidal disruption scheme \citep{fie11}. Most of our simulations are calculated on the $laohu$ GPU cluster in National Astronomical Observatories of China (NAOC). With up to million of particles in simulation, we can find out the dynamical evolution of stars near the recoiling SMBHs and estimate their tidal disruption rates, which may give us some hints for observation.

Our results show that the accretion of disrupted stars does not significantly impact the dynamical evolution of
recoiling SMBH. However, it changes the tidal disruption rate compared to a stationary SMBH system in a galactic center.
Most of tidal disruption events are contributed by unbound stars when the SMBH passing through the galactic center, and
the disruption rate in average is $\sim10^{-6} \accrate$, which is roughly an order of magnitude lower than the
stationary case, $\sim10^{-5} \accrate$. Besides, the tidal disruption rate for oscillating SMBH far from the galactic center
is roughly one or two orders of magnitude lower than stationary SMBH. We also find a bound stellar system around
the recoiling SMBH with a mass of $\sim 0.7\%$ black hole mass, which is consistent with \citet{mer09}. Except for
bound stellar systems, we find that there is a ``polarization cloud" composed of stars dynamically perturbed
by the oscillating SMBH. This ``polarization cloud" is axisymmetrically diffuse in a large spatial area which is
beyond the gravitational influence region of the oscillating SMBH. For this reason, most of the stars in the cloud are
not permanently bound to the SMBH. This kind of clouds are the echo of dynamical friction. The calculations for
parameter dependence show that the variation of initial mass of the recoiling SMBH can speed up or slow down the
dynamical evolution, and impact the tidal disruption rate. In addition, the variation of density slope can obviously
change the dynamical evolution and tidal disruption rate of the recoiling SMBH.

We give our galactic model and simulation method in Section \ref{method}. Some results about bound/unbound stellar
systems and tidal disruption rates for the stationary/recoiling SMBH are showed in Section \ref{results}. The dependence of
our results on other parameters are also investigated. Section \ref{dissum} gives discussion about the observational
implications and a short summary.

\section{GALACTIC MODEL AND $N$-BODY SIMULATIONS}
\label{method}

\subsection{Galactic Model}
\label{gala model}

For simplicity, we adopt a spherical Dehnen model to describe the ellipticals or bulges of galaxies \citep{deh93},
which is different from a S\'{e}rsic model used in \citet{gua08}. In a Dehnen model, the space density profile follows

\beq \rho(r)=\frac{3-\gamma}{4\pi}\frac{Ma}{r^\gamma(r+a)^{4-\gamma}} , \label{Dehnen rho} \eeq where $a$ is the scaling radius,
$M$ is the total mass of galaxies, and the slope index $\gamma$ is a constant between interval [0,3). The mass distribution is
proportional to $r^{-\gamma}$ for $r\ll a$ and $r^{-4}$ for $r\gg a$. In this model, it is easy for us to derive the galactic
potential, cumulative mass, and half-mass radius respectively \citep{deh93},

\beq \Phi(r) = \frac{GM}{a} \times \cases{ -\frac{1}{2-\gamma}\biggl[1-\biggl(\frac{r}{r+a}\biggr)^{2-\gamma}\biggr] & for $\gamma \neq 2$ \cr
\cr
\ln\frac{r}{r+a} & for $\gamma = 2$} , \label{eq:pot} \eeq

\beq M(r)=M\left(\frac{r}{r+a}\right)^{3-\gamma}, \label{eq:mass} \eeq

\beq
r_{1/2}=a \,[2^{1/(3-\gamma)}-1]^{-1}. \label{eq:halfr} \eeq where $G$ is the gravitational constant. From
Equation~(\ref{eq:pot}) the escape velocity at radius $r$ for $\gamma < 2$ is

\begin{eqnarray}
V_{\rm esc}(r) &=& \sqrt{\mid2\Phi(r)\mid} \nonumber \\
&=& \left(\frac{2}{2-\gamma}\frac{GM}{a}\right)^{1/2} \left|1-\left(\frac{r}{r+a}\right)^{2-\gamma}\right|^{1/2}, \, \rm{for} \, \gamma < 2.
\label{eq:vesc}
\end{eqnarray}
Thus the escape velocity from the galactic center for $\gamma < 2$ is

\beq V_{\rm esc}(0)=\left(\frac{2}{2-\gamma}\frac{GM}{a}\right)^{1/2} . \label{eq:vesc0} \eeq
If there is a SMBH with mass $\Mbh$
at the galactic center, we can estimate its influence radius $r_{\rm inf}$ by using the definition with stellar mass
$M_\star(r\leqslant r_{\rm inf})=2\Mbh$ as in \citet{mer06b}\footnote{This kind of definition is validated only for a
singular isothermal sphere nucleus. Here we adopt it also for $\gamma \neq 2$ to estimate $r_{\rm inf}$ at an order of magnitude.}

\beq
r_{\rm inf}=\frac{a\left(2\Mbh/M\right)^{\frac{1}{3-\gamma}}}{1-\left(2\Mbh/M\right)^{\frac{1}{3-\gamma}}}.
\label{eq:rinf}
\eeq

For simplicity, we adopt model units $G=M=a=1$ thereafter. In this new unit, the quantities relate to physical quantities with
the scale factor of time, velocity, and length, respectively

\begin{eqnarray}
    \label{eq:scalingT}
[T] &=& \left(\frac{GM}{a^3}\right)^{-1/2} \nonumber \\
    &=& 1.491\times 10^6(2^{\frac{1}{3-\gamma}}-1)^{3/2}\left( \frac{M}{10^{11}\msun}\right)^{-1/2}\left(\frac{r_{1/2}}{1\kpc}\right)^{3/2} \yr, \\
    \label{eq:scalingV}
[V] &=& \left(\frac{GM}{a}\right)^{1/2} \nonumber \\
    &=& 655.8\times (2^{\frac{1}{3-\gamma}}-1)^{-1/2}\left( \frac{M}{10^{11}\msun}\right)^{1/2}\left(\frac{r_{1/2}}{1\kpc}\right)^{-1/2} \kms, \\
    \label{eq:scalingr}
[R] &=& a=(2^{\frac{1}{3-\gamma}}-1)\left(\frac{r_{1/2}}{1\kpc}\right) \kpc.
\end{eqnarray}

With the model units, Equations~(\ref{eq:halfr}), (\ref{eq:vesc0}), and (\ref{eq:rinf}) can be rewritten

\beq
r_{1/2}=[2^{1/(3-\gamma)}-1]^{-1},
\label{eq:halfrD}
\eeq

\beq
V_{\rm esc}(0)=\left(\frac{2}{2-\gamma}\right)^{1/2},
\label{eq:vesc0D}
\eeq

\beq r_{\rm inf}=\frac{\left(2\Mbh\right)^{\frac{1}{3-\gamma}}}{1-\left(2\Mbh\right)^{\frac{1}{3-\gamma}}}\thickapprox
\left(2\Mbh\right)^{\frac{1}{3-\gamma}}.
\label{eq:rinfD}
\eeq

For a fiducial model adopted in the following calculations with $\gamma = 0.5$, $M = 4\times 10^{10} \msun$, $\Mbh =
4\times 10^{7} \msun$ and $r_{1/2} = 1 \kpc$, we can derive $V_{\rm esc}(0)\simeq 847\kms$ and $r_{\rm inf}\simeq
29\pc$.

\subsection{Tidal Disruption Scheme in $N$-Body Simulation}
\label{TD in simu}

Compared to a scattering experiment, direct $N$-body simulation can well represent the dynamical co-evolution of the
recoiling SMBH and surrounding stars, especially for the stellar interactions. However, it can not deal with
other process well, for example, the tidal disruption by SMBH. To solve this problem, we implement a special
tidal disruption scheme in our simulation.

A star with mass $m_*$ and radius $r_*$ will be tidally disrupted if it approaches a BH within the tidal radius \citep{hil75,ree88}

\begin{equation}
  r_{\rm t} \simeq \mu r_{*}(\Mbh/m_{*})^{1/3}.
  \label{eq:rt}
\end{equation}
Here $\mu$ is a dimensionless parameter of order unity. It is roughly $10^{-6}\, r_{\rm inf}-10^{-5}\,r_{\rm inf}$ for
the real conditions in galactic nuclei \citep{mer06b}. Based on Equation~(\ref{eq:rt}), to make sure that the tidal
radius is larger than the event horizon of the BH, there is a limitation for $\Mbh$. For a disrupted star with solar
radius $\rsun$ and mass $\msun$, a Schwarzschild BH should have $\Mbh \la 10^8 \msun$. This limitation mass can be
heavier when the BH has rapid spin.

In order to investigate the tidal disruption of the recoiling SMBH with $N$-body simulation, we have to find a proper way
to represent the tidal radius in our simulations. As shown in Equation~(\ref{eq:rt}), the tidal radius is proportional
to the radius of the disrupted star. However, in our $N$-body simulations without stellar evolution scheme,
stars are point like particles, whose radii are not defined. We can use the approximate scaling relation between
$r_{\rm inf}$ and $r_{\rm t}$ discussed above to set a proper value for $r_{\rm t}$, but this value is so small that
there will be only few tidal disruption events recorded in the simulation. That is because of the limitation on particle
resolution. Our simulation particle number is not enough even with $N=10^6$.

To solve this problem, a simplified strategy is adopted. We assume a larger $r_{\rm t}$ in our $N$-body simulation to
represent the tidal radius. The particle which get close to SMBH within $r_{\rm t}$ will be tidally disrupted and removed from the system with its mass added onto the SMBH. With this method, we can directly follow the
growth of $\Mbh$ and calculate the tidal disruption rate. However, the large $r_{\rm t}$ we adopted here will
overestimate the tidal disruption rate. The solution is to carry out a series of simulations with decreasing $r_{\rm
t}$ and extrapolate the results to the regime corresponding to real galaxies. Our extrapolation scheme is explained in
Section~\ref{calc num} in detail. We do not consider the growth of $r_{\rm t}$ for accreting SMBH or the momentum
transfer from tidal disrupted stars to the recoiling SMBH, because the total mass of disrupted stars is negligible compared
to the BH. A test simulation taking into account momentum transfer shows no significant difference in either the
dynamical evolution of recoiling SMBH or the stellar disruption rate.

\subsection{Numerical Method}
\label{num method}

To investigate the evolution of the recoiling SMBH with tidal disruption carefully, we make a series of integrations
with different parameters which are listed in Table \ref{tab:para}. The first column is the sequence number for
different models. Columns (2) - (4) give particle number $N$, tidal radius $r_{\rm t}$, and initial stellar density
slope $\gamma$ respectively. Column (5) is the initial mass ratio between $\Mbh$ and the total mass of system. Column (6)
gives initial recoil velocity in the unit of escape velocity at the galactic center.

All of our integrations adopt a parallel direct $N$-body $\varphi$\,-{\sc Grape}/$\varphi$\,-GPU code with fourth-order
Hermite integrator and simplified tidal disruption scheme. They all have time-step accuracy parameter $\eta=0.01$ and a
softening length $\epsilon=10^{-5}$. Most of our models adopt $N=10^6$ equal mass particles, with initial mass of SMBH
$\Mbh=0.001$ to represent the ellipticals/bulges and central SMBH respectively. To find out the dependency of the
results on particle number, several sets of integrations with $50\rm K$ and $250\rm K$ particles have been included.
There is also a calculation with $\Mbh=0.002$ to check the dependency on $\Mbh$. As mentioned before, for the
purposes of extrapolating to the smaller tidal radius in the real galactic center conditions, we have varied the $r_{\rm
t}$ from $5\times 10^{-5}$ to $5\times 10^{-2}$, which is roughly corresponding to $\sim 10^{-3}\,r_{\rm inf}$ to $\sim
r_{\rm inf}$ for $\gamma=0.5$. Since the density slope prior to the recoil is under debate, we choose $\gamma=0.5$
as our fiducial value, and also investigate other cases with $\gamma=1.0$ and $1.5$. The recoil velocity $V_{\rm k}$
is set to $V_{\rm k}=0.7V_{\rm esc}$ for most of the integrations, and there are calculations without recoiling
velocity for comparison. Besides, an integration with $V_{\rm k}=1.1V_{\rm esc}$ and $\gamma=1.0$ is included
for comparing with \citet{kom08b}.

Since Dehnen model provides the stellar distribution in a galaxy without central SMBH, we have to first to set up
consistent initial conditions for our system. At first, we put a SMBH into the center of a stellar system with Dehnen
profile, and evolve the entire system for 50 simulation time units to relax the core region. After that the system is
dynamically and thermally relaxed in relation to the SMBH gravitational potential. At this stage, we do not consider
stellar tidal disruption process. However, we want to make sure that stars inside the tidal disruption loss cone of the
central SMBH have already been removed by tidal accretion before the recoil of SMBH. Therefore, as a second step, we
switch on the tidal disruption scheme in the program and run it again with the SMBH still in equilibrium in the center
- this costs a few dynamical timescales at the influence radius. Thus all particles within the tidal disruption loss
cone can be removed before we apply a kick to the SMBH. After setting up the initial configuration of the stellar
distribution of the systems, we artificially give a recoil velocity to the SMBH and follow its dynamical evolution.

\section{RESULTS}
\label{results}

\subsection{Dynamical Evolution of the recoiling SMBH}
\label{Dyn evo}

\begin{figure}
\begin{center}
\includegraphics[width=0.9\textwidth,angle=0.]{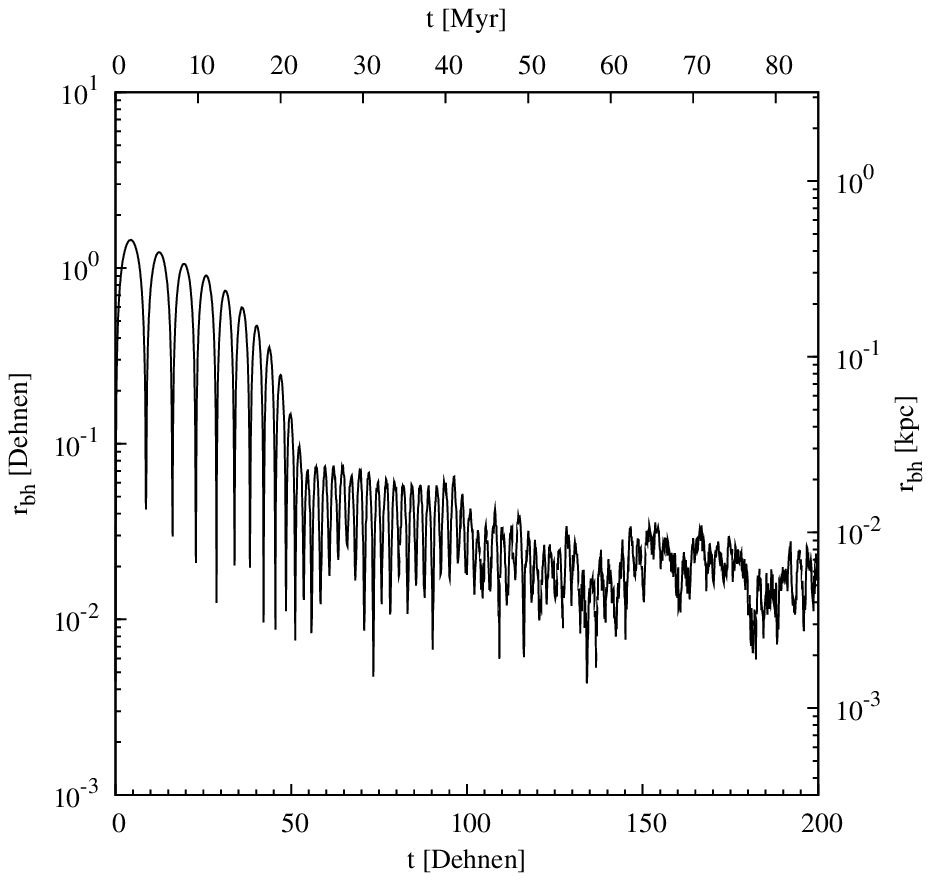}
\caption {Trajectory of the recoiling SMBH with model $06$. Here the bottom horizontal axis and the left vertical axis give the
evolved time and the displacement of BH $r_{\rm bh}$ respectively, with the units $G=M=a=1$. The right and the top axis
are transformed to real physical units with the assumption that $M = 4\times 10^{10} \msun$ and $r_{1/2} = 1 \kpc$. It
can be easily distinguished for the phase I and II which are similar to the result in \citet{gua08}. \label{fig:osc} }
\end{center}
\end{figure}

To carefully investigate the dynamical evolution of recoiling SMBH, we have a long time integration with model $06$ to
$t=200$, as shown in Figure~\ref{fig:osc}. It is clear that the evolution of the recoiling SMBH can be easily
divided into two different phases. The phase I has obviously damping trajectory from $t=0$ to $t\sim50$. After that,
the SMBH evolves into phase II, where the trajectory damped very slow for a long time. In phase I, the trajectory of
recoiling SMBH can be well predicted by Chandrasekhar's dynamical friction theory \citep{cha43}. However, this stage
will just keep till the SMBH's oscillation decayed to the size of core, where a slowly decayed phase II begins. That is
because the stellar mass interior to SMBH's orbit is roughly equal to $\Mbh$ at the end of phase I, where the
assumptions of Chandrasekhar's dynamical friction theory are invalid. We have tried to fit the trajectory analytically
in phase I with the same method as \citet{gua08} used, and got a similar result even though a Dehnen model is
adopted here instead of their core-S\'{e}rsic model.

If we continue the integration, the orbital oscillation of the SMBH will slowly decay and finally achieve to a thermal
equilibrium with surrounding stars, which is the so-called phase III in \citet{gua08}. Here we have not continue our
integrations into that stage because phase III is actually similar to the case of a stationary central SMBH, which is
out of the scope of this paper. We will just focus on phase I and phase II in the following integrations. Thus all of
our models integrate to $t=100$ in order to make sure that the evolution can achieve to phase II.

With regard to the orbital motion of the SMBH, our results agree qualitatively well with those of \citet{gua08}. That is
because the growth of the black hole due to tidal accretion, which did not take into account in \citet{gua08}, is
relatively small.

\subsection{Compact Star Clusters around the Recoiling SMBHs}
\label{stars}

A recoiling SMBH can carry a retinue of bound stars. \citet{mer09} predict that there will be a ``hypercompact stellar system"
(HCSS) bound to the recoiling SMBH. We can analytically estimate the bound population of the HCSS with Equations (1a) and (6a) in
\citet{mer09}. For example, the roughly boundary radius $r_{\rm ej}$ of HCSS should be

\begin{equation}
  r_{\rm ej} \equiv \frac{G\Mbh}{V_{\rm ej}^2},
  \label{eq:rej}
\end{equation}
with our Model $06$, where $V_{\rm ej}=0.7V_{\rm esc}\approx 0.8$, we have $r_{\rm ej}\approx 1.5625\times10^{-3}$ in our
simulation unit. Thus we can define a factor $f_{\rm b}$ to describe the fraction of bound stars,

\begin{equation}
  f_{\rm b} \equiv \frac{M_{\rm b}}{\Mbh}=F_{1}(\gamma)\left(\frac{r_{\rm ej}}{r_{\rm inf}}\right)^{3-\gamma},
  \label{eq:fb}
\end{equation}
where $M_{\rm b}$ is total mass of bound stars. Based on Equations ~(\ref{eq:rej}) and (\ref{eq:rinfD}), we have
$f_{\rm b}\approx 3.8845\times10^{-5}F_{1}(\gamma)$. For the case of $\gamma=0.5$, with $F_{1}(0.5)$ in Figure 1a of
\citet{mer09}, we have $f_{\rm b}\sim 0.001$. That means the ratio for HCSS mass to $\Mbh$ should be $\sim0.1\%$,
corresponding to just few particles for $1\rm M$ simulation. It should be aware that this fraction is just a lower limit
because our ``two times $\Mbh$" assumption actually overestimate $r_{\rm inf}$. To confirm the conclusion
above, we investigate this problem with $N$-body simulations.

In the simulation, it is very difficult to distinguish whether a star is bound to SMBH or not. That is because most of
star particles with negative total energies relative to the SMBH are loosely bound. Only few of them could orbit
the SMBH for more than one orbital period. For this reason, we identify a star to be bound to the recoiling SMBH if following two criteria are both satisfied: the particle (1) has negative total energy relative to the SMBH and (2) remains bound for at least one orbital oscillation period of the recoiling SMBH. We find that most of the ``bound" particles around SMBH particle will be evaporated before the end of simulation. This effect is more significant for our escaping SMBH calculation with model 10, where the number of bound particles decreases monotonically. For other oscillating SMBH calculations, there are many unbound particles can be captured by the SMBH. That means the interaction between the HCSS and stellar background is very important.

\begin{figure}
\begin{center}
\includegraphics[width=0.9\textwidth,angle=0.]{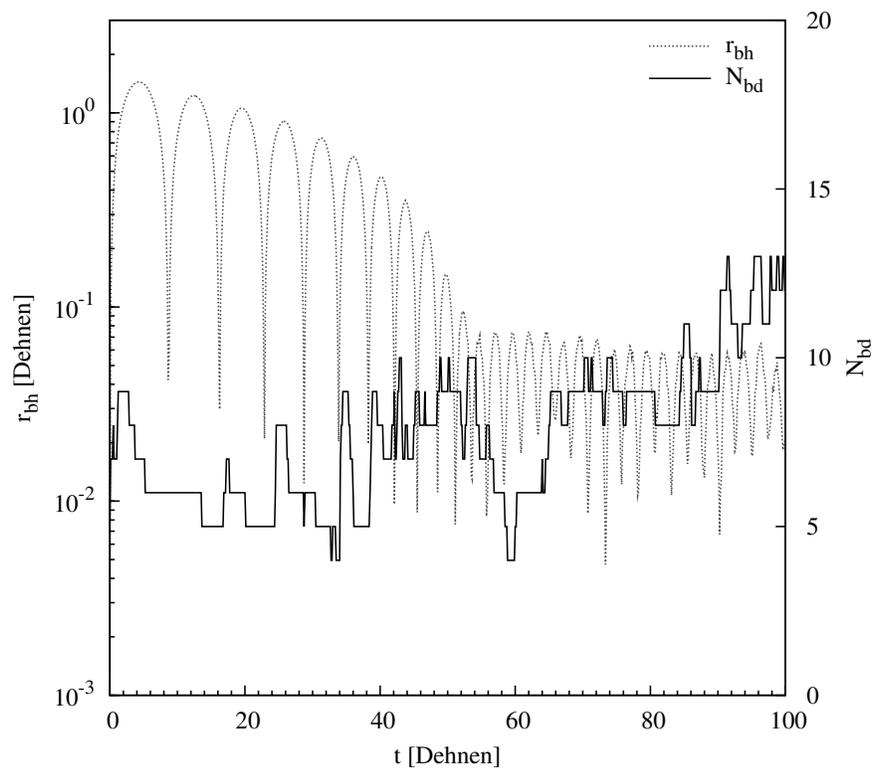}
\caption {The evolution of bound star numbers around a recoiling SMBH based on the model $06$. The horizontal and
vertical axis are the same as Figure~\ref{fig:traceAPO}, the right horizontal axis is the number count of the bound particles. The grey dotted line gives the trajectory of the recoiling SMBH, and the black solid line gives the variation of the bound particles. \label{fig:realbd} }
\end{center}
\end{figure}

Figure~\ref{fig:realbd} shows the evolution of bound stars around a recoiling SMBH in model $06$. At $t=0$, when
$V_{\rm ej}=0.7V_{\rm esc}$, there are five bound particles, roughly consistent with the estimation from
Equation~(\ref{eq:fb}). The number of bound particles, $N_{\rm {bd}}$, slightly increases during the evolution from
phase I to phase II. Because in phase II (1) the low relative velocity between stars and the recoiling SMBH increases
the probability of dynamical capture and (2) the stellar density around the SMBH in phase II is higher. Our results suggest
that a HCSS may grow via capturing unbound single stars. However, the low particle resolution and large statistical
error prevent us from further investigation of the growth of HCSS.

\begin{figure}
\begin{center}
\includegraphics[width=0.9\textwidth,angle=0.]{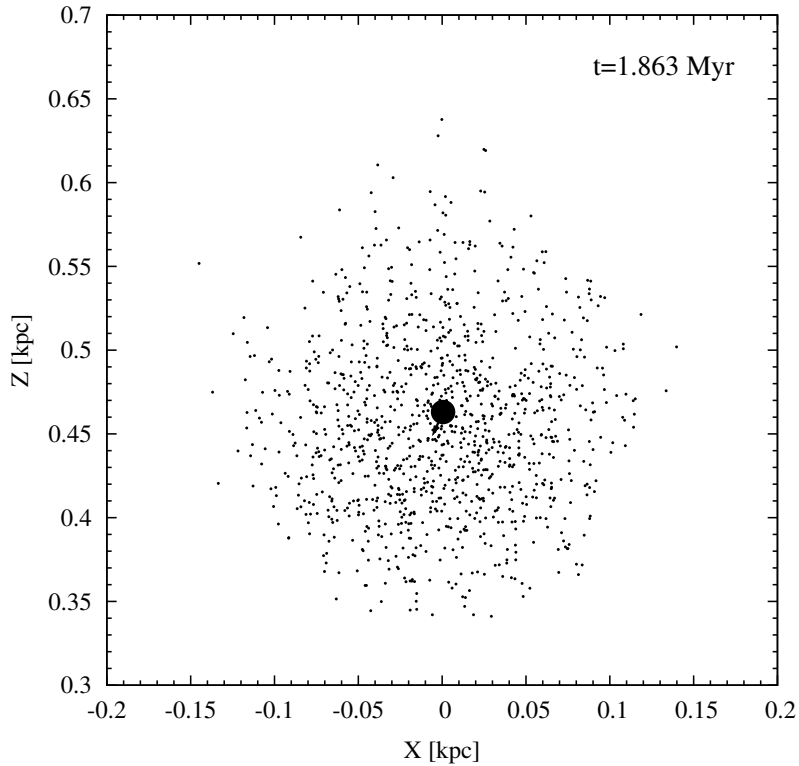}
\caption {Distribution of stars which are just strongly impacted by the recoiling SMBH at its first apocenter with model
$06$ for time snapshot $t=1.863\Myr$. We project the stellar distribution onto $X-Z$ plane, with the physical units under
the assumption that $M = 4\times 10^{10} \msun$ and $r_{1/2} = 1 \kpc$. The large black spot marks the position of
the recoiling SMBH. It can be seen that all of the impacted star particles form a cloud structure around the SMBH at that
moment. \label{fig:traceAPO} }
\end{center}
\end{figure}

\begin{figure}
\begin{center}
\includegraphics[width=0.9\textwidth,angle=0.]{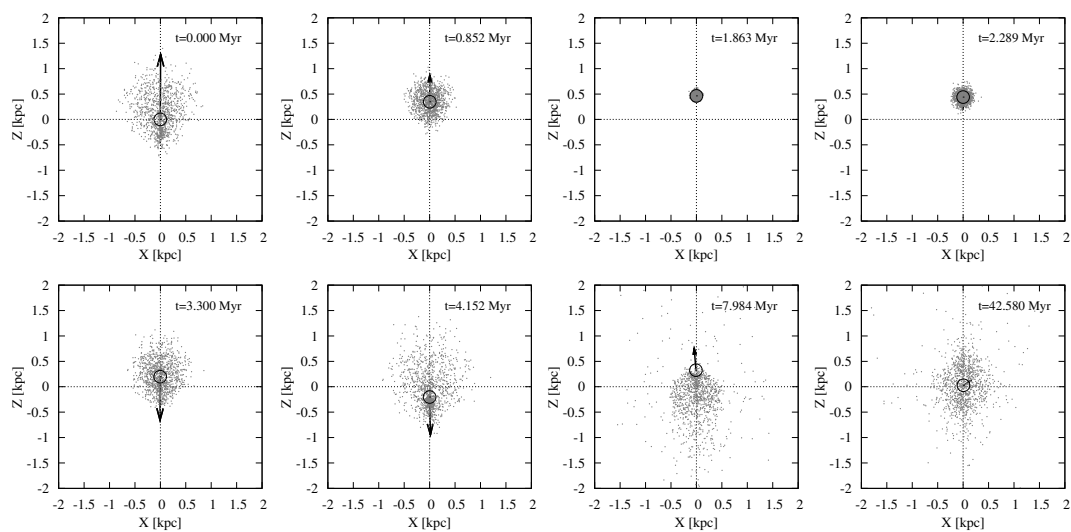}
\caption {Evolution of stars which are just strongly impacted by the recoiling SMBH at its first apocenter with model $06$. The
units and definition of axis are the same as Figure~\ref{fig:traceAPO}. The black circle and the grey dots mark the recoiling SMBH
and the stars respectively. The black arrow represents the size and orientation of BH's velocity. It can be seen that many unbound
stars are influenced by the SMBH. \label{fig:trace} }
\end{center}
\end{figure}

The HCSS may be detected as an off-nuclear compact system of similar size and stellar mass of a globular clusters but of having very high internal velocities \citep{mer09,ole11}. However, that is not the whole story. As shown in the Figures~\ref{fig:traceAPO} and~\ref{fig:trace}, we find that there is a stellar cloud composed of stars which are strongly impacted by the recoiling SMBH at
the first apocenter. The total mass of this cloud is comparable to $\Mbh$. They form a quasi-axisymmetric stellar cloud and always
have a dense region around SMBH. This kind of clouds gives an evidence for the echo of dynamical friction. For this reason,
we name that stellar cloud as ``polarization cloud". In our simulation of model 06, we choose a time snapshot when the SMBH arrives at its first apocenter, and pick up those polarization cloud members from the $E_{\rm tot}-r$ phase space which includes all of star
particles. All of polarization cloud members in the $E_{\rm tot}-r$ phase space are outliers around SMBH particle and can be easily
distinguished. Thus we can collect those polarization cloud particles and trace their evolution. However, it is very difficult to find an automatical solution to efficiently distinguish those polarization cloud members for every time snapshot. To deeply investigate this kind of polarization clouds, an efficient method to extract those outlier particles is needed. This problem should be solved in our following work.

In Figure~\ref{fig:traceAPO}, we plot that polarization cloud at
time $t=1.863\rm~ {Myr}$, when the recoiling SMBH arrives at its first
apocenter. All the parameters here are obtained from model $06$. In
order to investigate the origin and evolution of those cloud members, we
look back the data and follow the evolution of the stars from time $t=0$. Figure~\ref{fig:trace} gives the evolution of those
stars and the recoiling SMBH from $t=0$ until $t=42.6\rm~ {Myr}$,
when our computation stopped. The snapshots with $t=0\Myr,
0.852\Myr, 1.863\Myr$ in Figure~\ref{fig:trace} show the
evolutions of the cloud members from beginning toward the first apocenter
of the recoiling SMBH. The following three snapshots correspond to the
time when the recoiling SMBH comes back from the first apocenter
and passes through the density center for the second time. The
snapshot at $t=7.984\Myr$ shows the SMBH passing through galactic
center for the third time, and the last snapshot is for the end of
computation. Our calculation results and Figure~\ref{fig:trace} imply that the shape of polarization cloud varies with the oscillations of the recoiling SMBH. The comparison between the first and the last two panels vividly shows the changes of the cloud members
during the oscillation.

As shown in Figure~\ref{fig:realbd}, there are only small fraction of stars really bound to the SMBH, whereas the mass of
the polarization cloud is roughly equal to $\Mbh$. Therefore most of these cloud members are unbound to the recoiling
SMBH. After the recoil of SMBH, those polarization cloud members are accelerated and fall behind the oscillating SMBH.
During the oscillation, some of the stars gradually expand to larger area while others continuously follow the
SMBH to oscillate around the galactic center, and form an axisymmetrical distribution along the velocity
direction of the recoiling SMBH. After multiple interactions with oscillating SMBH, these stars obtain energy from
oscillating SMBH and diffuse to a large area. This is a vivid example that an oscillating SMBH transfer its orbital
energy to surround stars through dynamical friction.

In order to investigate the observational properties of this polarization cloud, we focus on its most compact stage,
when the recoiling SMBH arrive at the apocenter for the first time. The distribution of those stars are shown in
Figure~\ref{fig:traceAPO}. Here we scale the $N$-body system to our physical fiducial galaxy model with stellar mass $M
= 4\times 10^{10} \msun$ and half mass radius $r_{1/2} = 1 \kpc$. For such system, the polarization cloud has mass of
$M_{\rm csc}\simeq 10^{7} \msun$, and a size of diameter $D\simeq 260 \pc$. Further calculation shows that, its half
mass radius is $\sim 83 \pc$. This size and mass are much larger than a globular cluster and comparable to an
ultracompact dwarf galaxy. We also calculate the velocity dispersion of this polarization cloud. The line-of-sight
velocity dispersion along X, Y and Z axes are $185\kms$, $190\kms$ and $189\kms$ individually. That is significantly
higher than a typical ultracompact dwarf galaxy. Comparing with bound HCSS, this polarization cloud has larger size,
heavier mass and similar velocity dispersion. Since this cloud is mixed with HCSS, it may bring troubles to the
detection of HCSS. We also estimate the average surface density inside the half mass radius of the polarization cloud. The overdensity of the polarization cloud relative to the stellar background is only $\sim$ 4\%. This may be because of that the apocenter of the recoiling SMBH here is not very far from the galactic center, where the surface density of the stellar background is still relatively high in model 06 with $\gamma=0.5$. For a SMBH with a higher recoil velocity and a larger apocenter, the situation may be improved. To investigate this problem, more simulations are needed. We will investigate this problem in our future work.

It should be noted that our example above is just for one time snapshot of the evolving SMBH. Every snapshot could have
a similar stellar cloud.  The composition of these clouds may form an interesting structure which can trace the trajectory of the SMBH. The evolution and observation characters of this kind of structures will be discussed in our future work. For a recoiling SMBH with highly enough velocity to escape from the galaxy, that kind of polarization cloud may not be distinguishable because many
stars will be dropped behind the escaping SMBH. Without the multiple interactions between oscillating SMBH and
surrounding stars, the polarization cloud can not survive for a long time.

\subsection{Tidal Disruption}
\label{td}

\subsubsection{Calculate Tidal Disruption Rate Numerically} \label{calc num}

For a solar type star disrupted by a SMBH with mass $\Mbh=4\times 10^7 \msun$, based on Equation~(\ref{eq:rt}) and (\ref{eq:scalingr}), the tidal radius is $\sim 8\times10^{-6} \pc$, corresponding to $\sim 10^{-8}$ in simulation units with our fiducial model. As mentioned in Section \ref{gala model}, because of the limitation on particle
resolution, it is practically impossible with present computing capabilities to calculate the tidal disruption rate
directly (with a realistically small tidal radius) through direct $N$-body simulations.

To solve this problem, we perform a parameterized study, which treats both the particle number $N$ and the size of
the tidal disruption radius $r_{\rm t}$ as free parameters. All of the simulations have been done with these varying
parameters, and physical conclusions are drawn from extrapolating to the real galactic conditions. \citet{fie10} have
shown in Fokker-Planck models that such scaling is reliable and provides physically correct results. Besides, tidal
disruption events in our simulations, as well as the case in real galactic center, do not occur frequently, which means
that the increase of $\Mbh$ is discrete. Thus it is difficult to compute the tidal disruption rate through calculating
accreted masses in time bins. Instead, we measure the tidal disruption rate in a cumulative, averaged way - fitting the
mass growth of the black hole and derive its time derivative.

\begin{figure}
\begin{center}
\includegraphics[width=0.9\textwidth,angle=0.]{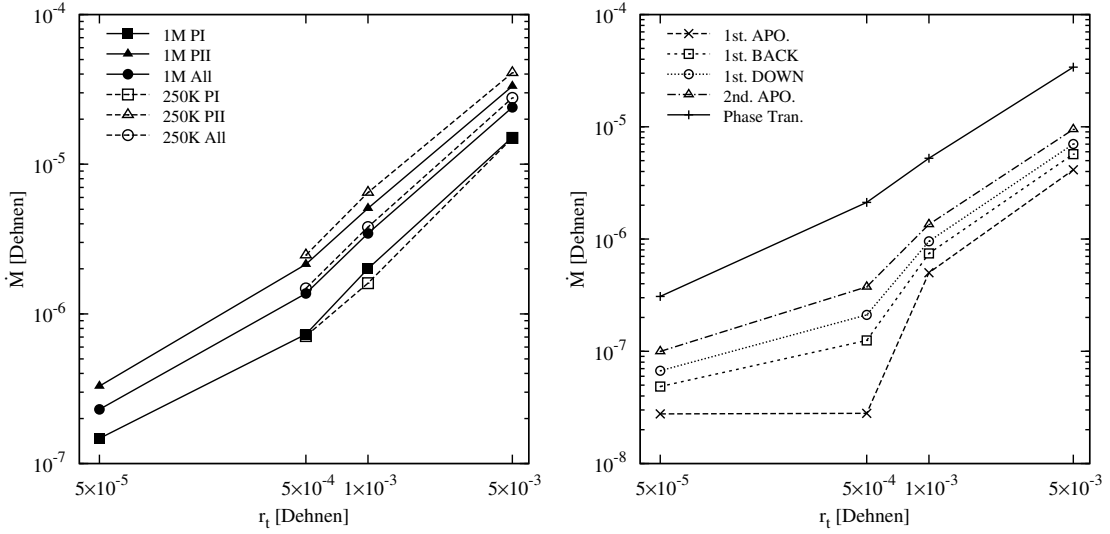}
\caption {Dependence of the tidal disruption rate on $N$ and $r_{\rm t}$. The left panel shows the dependence of the average
tidal disruption rates on $N$ and $r_{\rm t}$. The solid line and dashed line are the average disruption rates for $1\rm M$
and $250\rm K$ integrations respectively. The square, triangle, and circles respect to the average disruption rates for
phase I, phase II and the entire process respectively. The integrations are based on model $06$ with different particle
numbers and tidal radius. The integration for $r_{\rm t}=5\times 10^{-5}$ with $N=250\rm K$ is not included because the
amount of resulted tidal disruption events is too small to calculate an average rate. The right panel shows the disruption rate
- $r_{\rm t}$ dependence for several key points with integrations for $N=1\rm M$. Here ``1st. APO.", ``1st. BACK",
``1st. DOWN", ``2nd APO." and ``Phase Tran." are corresponding to the moment which SMBH the first time arrive at its
apocenter, the first time come back to the density center, the first time arrive at apocenter on the other side, the second time
arrive at its apocenter and the transition from phase I to phase II respectively. For both of two panels, horizontal and
vertical axis give the tidal radius and tidal disruption rate in the units of $G=M=a=1$ respectively.\label{fig:TDNRT}
}
\end{center}
\end{figure}

In order to find out the extrapolation relation, the dependence of tidal disruption rate on particle numbers and tidal radius
should be investigated. Figure~\ref{fig:TDNRT} shows the disruption rates changes with different $N$ and $r_{\rm t}$, with
parameters are $N=250\rm K, 1\rm M$, $r_{\rm t}=5\times 10^{-3}, 1\times 10^{-3}, 5\times 10^{-4}, 5\times 10^{-5}$,
$\gamma=0.5$, $\Mbh=0.001$ and $V_{\rm ej}=0.7V_{\rm esc}$. The left panel shows that the dependence of averaged tidal disruption
rates on $N$ and $r_{\rm t}$. Here we calculate the averaged disruption rates for phase I, phase II and entire process separately
with different $N$ and $r_{\rm t}$. These integrations show that the dependence of tidal disruption rate on particle number is
very weak. Further discussions are showed in Section \ref{Nd}.

Based on the loss cone feeding theory \citep{fra76,lig77}, contrary to empty loss cone case, the tidal disruption rate
for full loss cone does not depend on $N$. It should be notice that the tidal disruption rate here is for mass
disruption rate, and the disrupted particles indeed depend on $N$. The slight different average rates between $1\rm M$
and $250\rm K$ cases means that the loss cones in our integrations are nearly full and the changes of particle numbers
do not bring significant impact.

Opposed to particle numbers, the changes on $r_{\rm t}$ strongly influence the disruption rates. A detailed study is in
the right panel of Figure~\ref{fig:TDNRT}, which shows the evolution of disruption rate - $r_{\rm t}$ dependence for
several special points with integrations for $N=1\rm M$. Five special snapshot have been picked out. Here ``1st. APO.", ``1st. BACK", ``1st. DOWN", ``2nd APO." and ``Phase Tran." refer respectively to the snapshot that the oscillating SMBH the first time arrives at the apocenter, the first time returns to the density center, the first time reaches the opposite apocenter, the second time return to the apocenter and the time of transition from phase I to phase II. Our results show that the disruption rates increase for every $r_{\rm t}$ value during the evolution of the recoiling SMBH. That is because a decayed orbit of SMBH leads to a growing stellar background around it. Besides, the similar linear disruption rate - $r_{\rm t}$ relation as left panel has been found. However, the relation here for the first four snapshots are not as good as the average results in the left panel. That may because our particle resolution for small $r_{\rm t}$ is not good enough in phase I.

Based on our result data, it seems that there is an approximate linear relation between $r_{\rm t}$ and disruption rates. As a
result of full loss cone condition, the disruption rates will be proportional to cross section $S$ and density $\rho(r)$ near the
recoiling SMBH,

\beq
  \dot{M}\simeq S\rho(r) \overline{v}_{\rm rela},
  \label{eq:tdr}
\eeq where $\overline{v}_{\rm rela}$ is average relative velocity between the recoiling SMBH and the surrounding stars.

It is difficult for us to analytically derive stellar distribution around a recoiling SMBH. However, the cross section can be
wrote as

\beq
  S\simeq\pi r_{\rm t}^2\left(1+\frac{2G\Mbh}{r_{\rm t}\overline{v}_{\rm rela}^{2}}\right),
  \label{eq:CS}
\eeq which can write the tidal disruption rate as

\beq
  \dot{M}\simeq\pi r_{\rm t}^2\rho(r) \overline{v}_{\rm rela}\left(1+\frac{2G\Mbh}{r_{\rm t}\overline{v}_{\rm rela}^{2}}\right).
  \label{eq:mdot}
\eeq For $r\gg r_{\rm t}$ that is $2G\Mbh/r_{\rm t} \gg \overline{v}_{\rm rela}^{2}$ (even if one takes into account that
$\overline{v}_{\rm rela}^{2}$ contains a contribution from the SMBH kick velocity), so we are in the gravitational
focusing regime, where

\beq
  S\simeq \frac{2\pi G\Mbh}{\overline{v}_{\rm rela}^{2}}r_{\rm t}.
  \label{eq:CSp}
\eeq
Thus we have

\beq
  \dot{M}\simeq \frac{2\pi G\Mbh}{\overline{v}_{\rm rela}}\rho(r)r_{\rm t}.
  \label{eq:mdotsimp}
\eeq

If $\rho$ and $\overline{v}_{\rm rela}$ are constant with time, the tidal accretion rate will scale linearly with
$r_{\rm t}$. This relation is clearly only correct in a time averaged sense, because when the SMBH oscillating in the
galaxy, all quantities will be strongly changed within a single orbit (density, stellar velocity dispersion and SMBH
velocity, the two latter quantities determining $\overline{v}_{\rm rela}$). However, in a running time averaged over
several orbits, our results show that there is not a large variation in these quantities. So we can define a
dimensionless tidal accretion parameter $\alpha$ by the following relation:

\beq
  t_{\rm dyn} \frac{\dot{M}}{\Mbh} = \alpha \frac{r_{\rm t}}{\overline{v}_{\rm rela} t_{\rm dyn}} .
  \label{eq:alpha}
\eeq Here $t_{\rm dyn}$ is the dynamic time scale around the position of the recoiling SMBH. Our simulations imply that
across all of our model families $\alpha$ does not depend strongly on $N$ or $r_{\rm t}$, but only on the evolutionary
phase.  Thus $\alpha$ can be used in extrapolations to the real system with very small $r_{\rm t}$ and large $N$. Based
on the analysis above, though it is complicated to get an accurate scaling relation, we still can estimate the tidal
disruption rate through linear extrapolation of tidal radius.

\subsubsection{Tidal Disruption Rate for the Stationary SMBH} \label{tdrs}

\begin{figure}
\begin{center}
\includegraphics[width=0.9\textwidth,angle=0.]{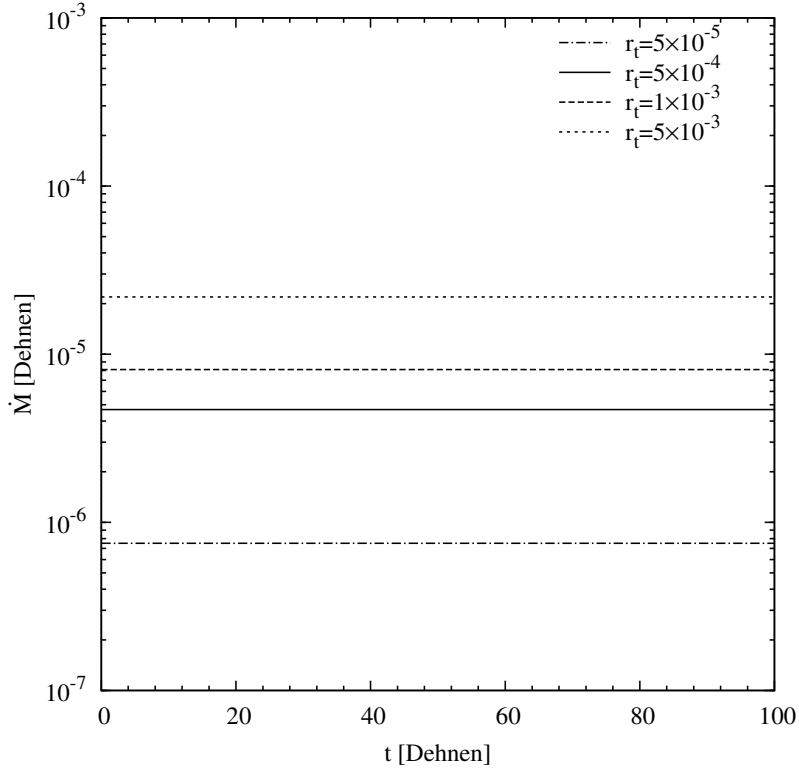}
\caption {Tidal disruption rates for the stationary SMBH with different $r_{\rm t}$. Here the horizontal and vertical axis give the
evolved time and tidal disruption rate in the units of $G=M=a=1$ respectively. The dotted, dashed, solid and dash
dotted lines are calculations with different tidal radii $r_{\rm t}=5\times 10^{-3}, 1\times 10^{-3}, 5\times 10^{-4},
5\times 10^{-5}$, which correspond to model $15,17,05$ and $19$ respectively. \label{fig:TDR} }
\end{center}
\end{figure}

For comparison purposes, we calculate the tidal disruption rate of a stationary SMBH without recoil in the galactic center.
A series of simulations with different $r_{\rm t}$ values are carried out. Figure~\ref{fig:TDR} shows the static
disruption rate for $r_{\rm t}=5\times 10^{-3}, 1\times 10^{-3}, 5\times 10^{-4}, 5\times 10^{-5}$, corresponding
to model $15$, $17$, $05$ and $19$ respectively. As mentioned in Section \ref{calc num}, the tidal radius for our fiducial model is $\sim 8\times 10^{-6} \pc$, corresponding to $\sim 10^{-8}$ in our simulation units. Our simulation results with different particle numbers for stationary case show that the tidal disruption rates weakly depend on the particle numbers. Thus we can loosely accept a full loss cone condition for the stationary case. As discussed in Section \ref{calc num}, there is a roughly linear correlation between tidal disruption rate and $r_{\rm t}$ for full loss cone case. Based on Equation~(\ref{eq:mdotsimp}), we can extrapolate our simulation results to the real galaxy conditions with smaller tidal radii. Figure~\ref{fig:TDR} shows that the tidal disruption rate for $r_{\rm t}=10^{-3}$ is $\sim 8\times 10^{-6}$ in our simulation units. That can be extrapolated to $r_{\rm t}\sim 10^{-8}$, and gives us the tidal disruption rate $\sim 10^{-10}$, which corresponds to the order of magnitude $\sim10^{-5} \accrate$ for our fiducial model.

\subsubsection{Tidal Disruption Rate for the recoiling SMBH} \label{tdrr}

In a recoiling SMBH system, there are two populations of stars contributed to the tidal disruption rate. One is the bound
stars around the SMBH, and the other is those unbound stars encountered with the recoiling SMBH. If the bound stars dominate, as
demonstrated in \citet{kom08b} for an escaping SMBH, the tidal disruption rate should roughly keep a constant level at
the beginning, and finally drop down because of the consumption of bound stars. On the contrary, if the tidal
disruption rate is dominated by unbound stars, the results will depend on stellar density, relative velocity of SMBH,
and the size of tidal radius, which described by  Equation~(\ref{eq:mdot}).

If we just consider about the contribution of the unbound stars, for an oscillating SMBH, its tidal disruption
rate near the galactic center should be higher than the apocenter region because of the different stellar density. However, $\overline{v}_{\rm rela}$ can also change the result. For instance, we have an estimation in our Model $16$
with $r_{\rm t}=5\times 10^{-3}$, $\gamma=0.5$, $\Mbh=0.001$ and $V_{\rm ej}=0.7V_{\rm esc}\simeq 0.8$. Our simulation
result shows that the first apocenter of the recoiling SMBH is around $r_{\rm BH}\sim 1.4$. With the assumptions that
$\overline{v}_{\rm rela}(r=1.4)\sim 0.1$ and $\overline{v}_{\rm rela}(r=0)\sim 0.8$, we can estimate the disruption
rate ratio from the galactic center to the first apocenter region. By substituted into Equations~(\ref{Dehnen rho}) and
(\ref{eq:mdot}), that ratio is $\sim 8$. When scaling to the real galactic center conditions with $r_{\rm t}\sim 10^{-8}$, it
should be $\sim 3$. Moreover, this ratio also depends on $\overline{v}_{\rm rela}$. If $\overline{v}_{\rm rela}$ near apocenters
can be larger than $0.1$, that ratio may be higher. Unfortunately, we can not obtain an accurate value for
$\overline{v}_{\rm rela}$. For estimation, it can be seen that the tidal disruption rates near the galactic center are
several times or even an order of magnitude higher than the apocenter region.

\begin{figure}
\begin{center}
\includegraphics[width=0.9\textwidth,angle=0.]{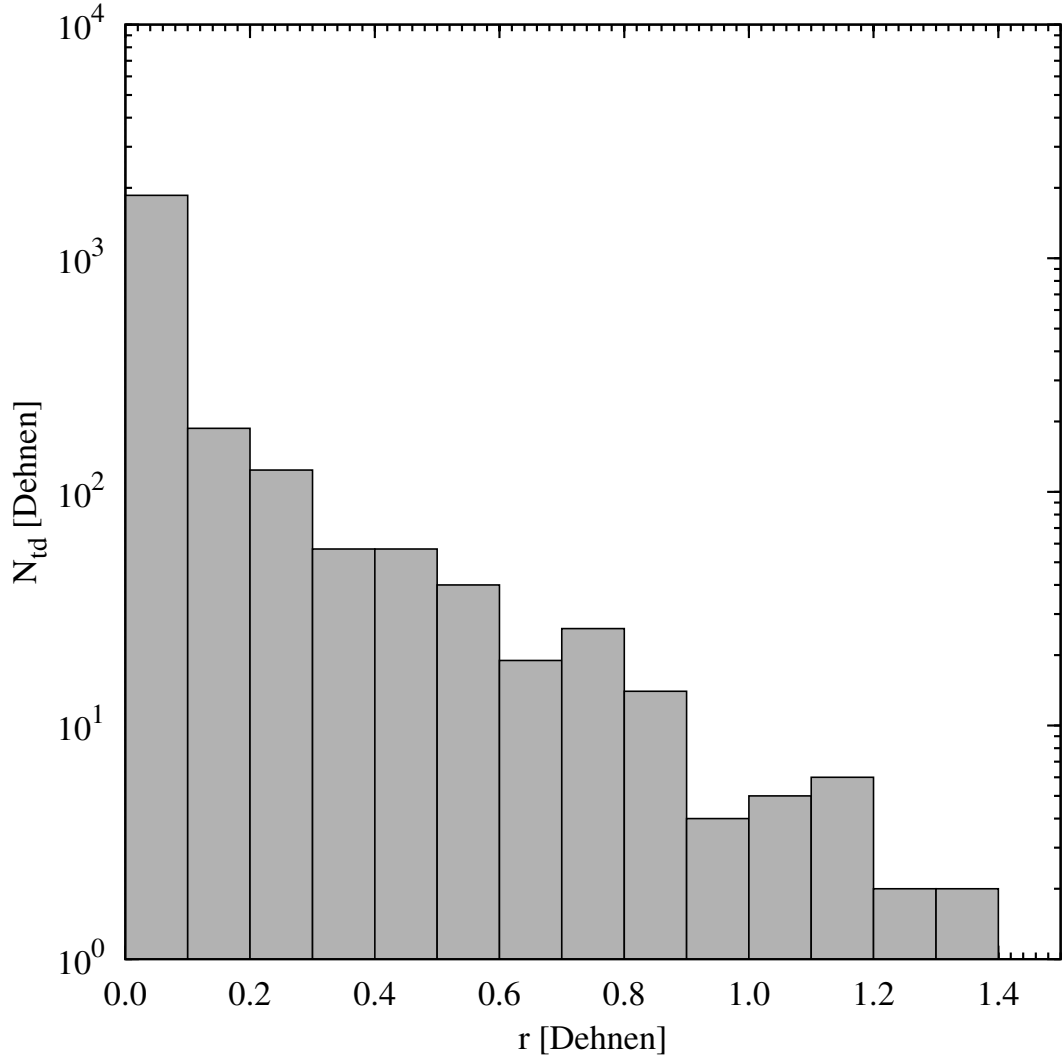}
\caption {The distribution of the cumulative tidal disruption events count $N_{\rm td}$ relative to $r$ with model $16$.
Here the horizontal and vertical axis give the cumulative tidal disruption events count and $r$ respectively. The shaded histogram
marks the cumulative tidal disruption events count for the different distance. \label{fig:TDE} }
\end{center}
\end{figure}

That conclusion is confirmed by our simulation results. Figure~\ref{fig:TDE} gives the distribution of the tidal disruption
events count $N_{\rm td}$ relative to distance $r$ in model $16$. It shows that most of the tidal disruption
events appear around the density center, which indicates a boosted tidal disruption rate when the oscillating SMBH pass
through the galactic center. A SMBH around the apocenter with a nearly zero velocity always corresponds to a low
$\overline{v}_{\rm rela}$, which should give us a high disruption rate. However, as Equation~(\ref{eq:mdot})
shown, the low stellar density environment makes the tidal disruption events less than at galactic center. The higher
disruption rate near the galactic center means that, for an oscillating SMBH, most of tidal disruption events are
contributed by unbound stars. For this reason, unbound stars are very important for calculating tidal disruption rate.

\begin{figure}
\begin{center}
\includegraphics[width=0.9\textwidth,angle=0.]{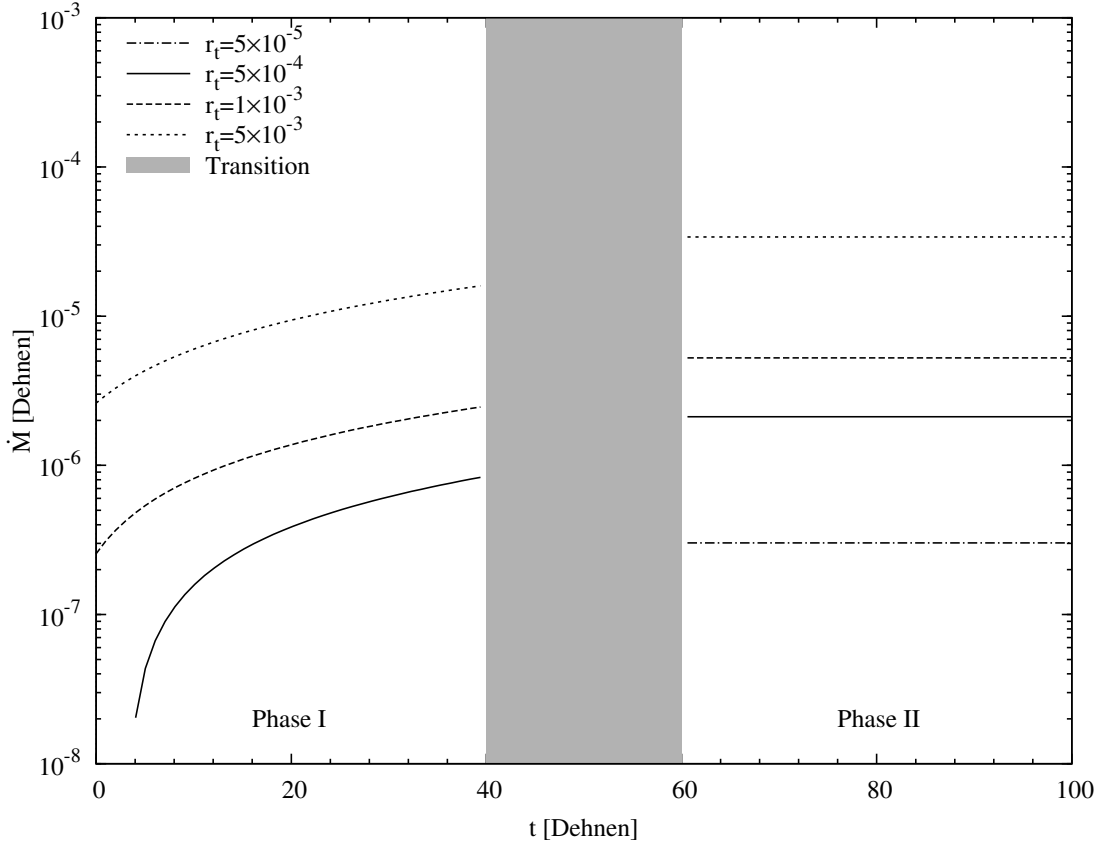}
\caption {The time evolution of the tidal disruption rates with different $r_{\rm t}$. The left part and right part are
the fitted results for phase I and phase II respectively. The shaded region corresponds to a transition area from phase I to
phase II. Axis are labeled as same as Figure~\ref{fig:TDR}. The dotted, dashed, solid and dash dotted lines are the
calculations with different tidal radii $r_{\rm t}=5\times 10^{-3}, 1\times 10^{-3}, 5\times 10^{-4}, 5\times 10^{-5}$,
corresponding to model $16,18,06$ and $20$ respectively. Since the model $20$ with $r_{\rm t}=5\times 10^{-5}$ does not
have enough particle resolution to calculate the disruption rate during phase I, it is absent in the left part.
\label{fig:TDRK} }
\end{center}
\end{figure}

Figure~\ref{fig:TDRK} shows the tidal disruption rates for a recoiling SMBH with different $r_{\rm t}=5\times 10^{-3},
1\times 10^{-3}, 5\times 10^{-4}, 5\times 10^{-5}$, which belong to model $16,18,06$ and $20$ respectively. To increase
the accuracy of our fitting results for disruption rate in phase I and phase II, we neglect the transition boundary
between the two phases in the fitting. Therefore, we fit disruption rate for phase I and phase II separately. As a
result, the fitted results are not continuous between two phases.

The left part of Figure~\ref{fig:TDRK} shows that the tidal disruption rates in phase I are linearly
increasing. For different $r_{\rm t}$ values, there is also a similar linear relation to stationary SMBH case, even
though there is an increasing tidal disruption rate instead of a constant value. Here the calculation with
model $20$ ($r_{\rm t}=5\times 10^{-5}$) is not included because the particle resolution is too low to estimate
disruption rate. With the same extrapolating relation described in Section~\ref{calc num}, we estimate that the averaged tidal
disruption rate for phase I should around the order of magnitude $\sim10^{-6} \accrate$, which is about an order of
magnitude lower than stationary SMBH in the galactic center.

For the phase II, when the orbit of the recoiling SMBH damped to relative small region, the variations of stellar density
and velocity of SMBH are smaller than phase I. Thus the difference of disruption rate between apocenters and galactic
center is not so intensive. Since the tidal disruption events during entire phase I are few in our simulation, the
accuracy for calculating tidal disruption rate is not very good. However, there are plenty of tidal disruption events
in phase II, which provides a better accuracy to calculate the tidal disruption rate in phase II. Actually, as
shown in the right part of Figure~\ref{fig:TDRK}, there are constant disruption rates as similar as the stationary
case. That is because (1) the stellar densities around SMBHs in phase II and in the stationary SMBH case are similar and
(2) their loss cones are both full.

\subsection{Dependence on Particle Number} \label{Nd}

\begin{figure}
\begin{center}
\includegraphics[width=1.0\textwidth,angle=0.]{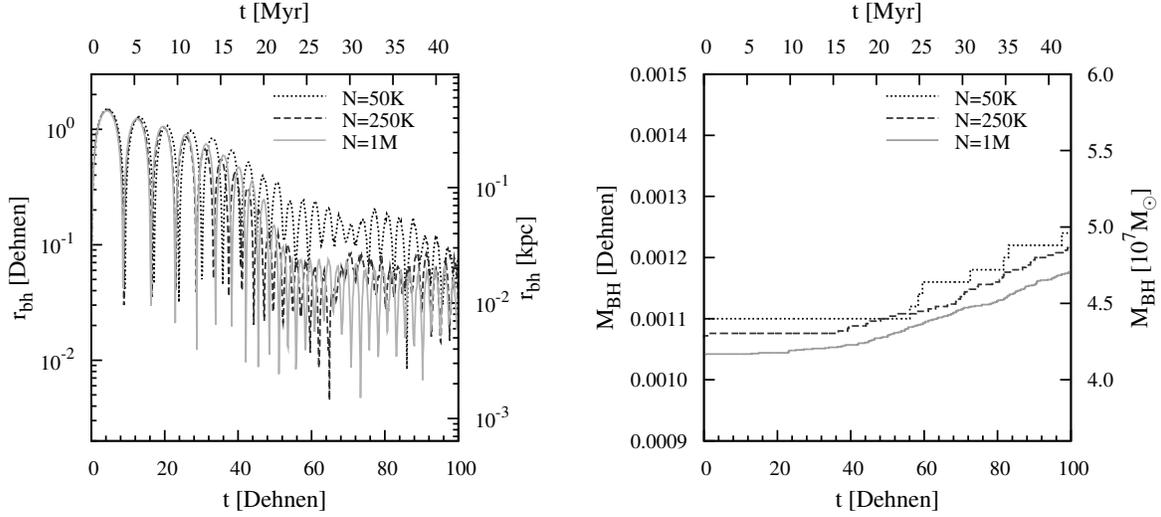}
\caption {The particle number dependence for the recoiling SMBH orbital oscillation and the mass increasing with model $01$, $02$
and $06$. Axis are labeled as same as in Figure~\ref{fig:osc}. The dotted, dashed and solid lines are the calculations
with particle numbers $N=50\rm K, 250\rm K, 1\rm M$ respectively. Left panel is the damped orbital evolutions of recoiling
SMBHs with various $N$. It can be seen that the $250\rm K$ and $1\rm M$ cases have a similar oscillate amplitude in phase
II, while the $50\rm K$ is different from others because of the poor particle resolution. The right panel shows the mass
increasing for the recoiling SMBH. Here the vertical axis represents to the mass of the recoiling SMBH. The different initial
mass of the recoiling SMBH is due to the special relaxation scheme described in Section~\ref{num method}. Only the
calculations with $250\rm K$ and $1\rm M$ particles can give us relatively smooth mass increasing curve, and all the
calculations achieve to similar order of magnitude for final mass. \label{fig:Ndep} }
\end{center}
\end{figure}

Since $1\rm M$ particles do not have enough high resolution to represent a real galaxy, it is important for us to find out the
dependence of our results on the particle number. Only with this considered, the direct $N$-body simulation results for dynamical
properties of the recoiling SMBHs and surrounding stars can be extrapolated to real conditions in galactic nuclei. To bring this
forth, a series of integrations with various $N$ and same other parameters have been done, which relative to models $01$, $02$
and $06$ with $N=50\rm K,250\rm K,1\rm M$ respectively. Figure~\ref{fig:Ndep} shows the particle number dependence of
the recoiling SMBH orbital oscillation and mass increasing for model $01$, $02$ and $06$ respectively. It can be seen that the
$250\rm K$ and $1\rm M$ cases have similar oscillate amplitudes in phase II, whereas the $50\rm K$ run can not well represent the
damped evolution of the recoiling SMBH. Besides, only the calculations with $250\rm K$ and $1\rm M$ particles can give relatively
smooth mass increasing curve, and they all achieve to the similar final mass.

Limited by the particle resolution, we can not obtain good enough mass evolution data to fit the disruption
rates in phase I both for $50\rm K$ and $250\rm K$ integrations with model $01$ and $02$. The only thing can be done is
to calculate average tidal disruption rates for phase I and II, as discussed in Section~\ref{td} with
Figure~\ref{fig:TDNRT}. However, those problems do not exist for all of $1\rm M$ integrations and other $250\rm K$
integrations. For this reason, most of our calculations adopt $1\rm M$ particles. As shown in Figure~\ref{fig:TDNRT},
the difference of average disruption rates between $250\rm K$ and $1\rm M$ is very small, which also indicates that the
loss cones are nearly full.

With $1\rm M$ particles simulation, the mass ratio of SMBH to stars should be $\sim10^3$. This ratio is smaller
than the real galaxy condition which should be $\sim10^6-10^9$. The small mass ratio between SMBH and stars may lead to a
higher Brownian velocity comparing with real galaxy condition \citep{mer07}. However, most of our simulations are
terminated before phase III, which makes the amplitude of oscillating SMBH is greater than Brownian amplitude. Detailed
discussion in \citet{gua08} with series of calculations for different particle numbers shows that the particle mass can
not significantly impact the dynamical evolution for both phase I and II.

The limitation of particle number can also influnece the relaxation time $T_{\rm r}$. For a homogenous isotropic
distribution with equal-mass stars, the two-body relaxation time in our simulation units is \citep{spi87}
\beq
T_{\rm{r}}(r)\approx\frac{0.34\sigma^3}{\rho(r) m_{*} \ln{\Lambda}} = \frac{0.34N\sigma^3}{\rho(r)\ln{\Lambda}},
\label{eq:tr}
\eeq
where $\ln{\Lambda}$ is Coulomb logarithm and we can estimate it as
\beq
\ln{\Lambda}\approx\ln\frac{r_{\rm inf}\sigma^2}{2 m_{*}} = \ln\frac{Nr_{\rm inf}\sigma^2}{2}.
\label{eq:cl}
\eeq

When $N$ is large enough, for instance $N \gtrsim 10^6$, the relaxation time will be roughly proportional to $N/\ln{N}$.
It means that our $N$-body simulation with $1\rm M$ particles always give us a shorter relaxation time comparing with
real galaxy condition. This fast relaxation in $N$-body simulation may bring troubles to our results. If the core
region in our simulation has relaxed before the recoiling SMBH return, our results will departure from the real galaxy
condition, because the relaxation time in a real galactic core is always very long.

To solve this problem, we calculate the relaxation time of the central core region for both the simulation model
and real galaxy condition. We choose $r=r_{\rm inf}$, $N=10^6$ and $N=4\times10^{10}$ in model $06$ to calculate the
relaxation time $T_{\rm{r}}(r_{\rm inf})$. Since we know little about the exactly value of velocity dispersion $\sigma$
around $r=r_{\rm inf}$, we range the $\sigma$ from $10^{-2}$ to 1. For the real galaxy condition, our estimations show that, the $T_{\rm{r}}(r_{\rm inf})$ is always longer than the half period of the oscillating SMBH. That means the core region with a real
galaxy condition can not relax before the SMBH return. The same result applied to our simulations with
$\sigma\gtrsim0.06$ for $N=10^6$. Based on Equation~(\ref{eq:scalingV}), $0.06$ corresponds to $\sim45\kms$, which is
smaller than most of the galactic center cases. Thus we can conclude that the core region can not relax before the
recoiling SMBH return both for simulation and real galaxy condition.

The limitation of particle number in $N$-body simulation may also impacts the results for the stationary SMBH. In recoil
case, the oscillating SMBH always corresponds to the full loss cone status both for the simulation and the real galaxy condition.
For a stationary SMBH, things will be different. Both the analytical calculations and the numerical simulations for long term evolution
of tidal disruption from stationary SMBH show that the loss cone in real galaxy is empty for most of the cases
\citep{fra76,lig77,coh78,bau04}. However, our calculations are just focus on a relative short period, which correspond
to $\sim 10^7 \yr$ for our fiducial model. During this short period, the loss cone could be roughly full. This
prediction has been confirmed by our simulation results.

Based on discussions above, there are some conclusions: (1) both $250\rm K$ and $1\rm M$ integrations are good enough
to represent dynamical evolution of the recoiling SMBH, (2) in order to investigate the tidal disruption and co-evolution
of the recoiling SMBH with surrounding stars, it is better to use integrations with $N=1\rm M$, (3) the small
difference on average disruption rates between $250\rm K$ and $1\rm M$ integrations indicates that the loss cones are
nearly full, otherwise the results should change obviously with various particle numbers, (4) the particle number we
adopt in simulation is less than the real galaxy condition, which can lead to a shorter relaxation time scale. However,
it can not significantly impact our results.

\subsection{Dependence on other Parameters} \label{depence}

Based on the Chandrasekhar's dynamical friction theory \citep{cha43}, the dynamical friction time scale is inversely
proportional to the mass of the recoiling SMBH. For this reason, a heavier SMBH should has shorter evolution time scale in phase
I. In order to investigate the dependence of evolution time scale on black hole mass $\Mbh$, we run a test model $07$
with $\Mbh = 0.002$. Comparing with model $06$, we find that model $07$ gives a shorter oscillation period in phase I,
which is roughly two times smaller than that in model $06$. That is consistent with the prediction of dynamical
friction theory. Besides, the tidal disruption rate in model $07$ is roughly two times higher than that in model $06$
for both of the two phases, which is roughly consistent with Equation~(\ref{eq:mdotsimp}). For a galaxy with lighter
SMBH, which has $\Mbh < 10^{-3}M$, there will be a longer dynamical evolution time in phase I and smaller tidal
disruption rate comparing with our results above. Limited by the calculation capability, simulations with the lower mass SMBH
will cost so much time that we can not investigate them carefully.

We note that in the simulations with very large tidal radius, for example, $r_{\rm t} = 0.05$ in model $13$ and $14$,
rapid increase on $\Mbh$ lead to the declined disruption rates. This significant growth of $\Mbh$ is artificial, which makes the declination of disruption rate is unphysical.

When considering different density slopes, the results are quite different. The higher of $\gamma$ value is, the faster of the
system evolved. Our integration with model $12$ shows that the recoiling SMBH will enter phase II around $t=5$, which is more
faster than evolution of SMBH in model $06$ with $t\sim 50$. Thus our models for $\gamma=1.0$ and $1.5$ actually reach phase III
at the end of integration. That can be easily understood because a larger $\gamma$ means a denser cusp in the galactic center.
For the same reason, a larger $\gamma$ also leads to a higher disruption rate.

\begin{figure}
\begin{center}
\includegraphics[width=0.9\textwidth,angle=0.]{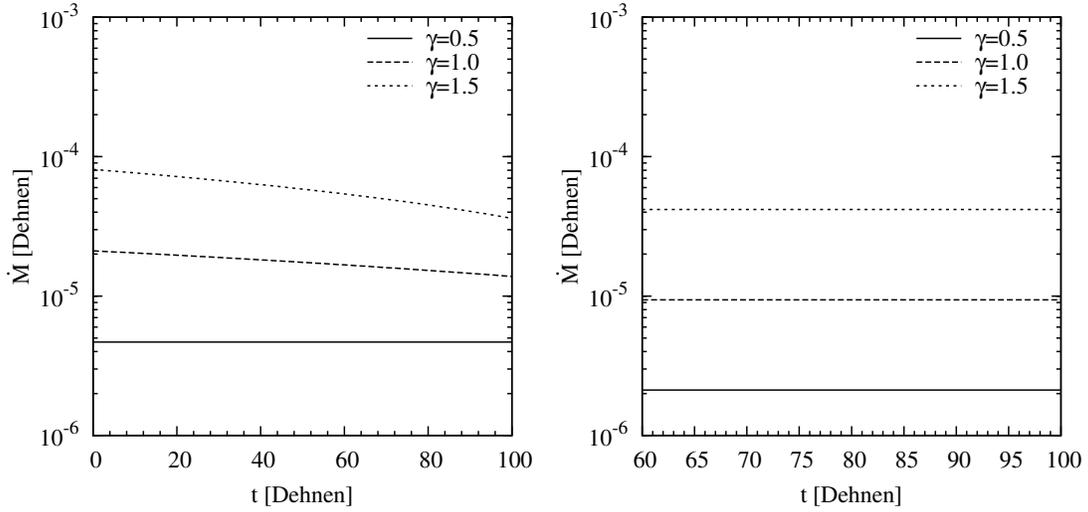}
\caption {The tidal disruption rates for different $\gamma$ values and different models. The left panel is the disruption rates
of stationary SMBHs with model $05$, $08$ and $11$. The right panel shows disruption rates of the recoiling SMBHs for the time
interval $t=60-100$, in model $06$, $09$ and $12$. Here the horizontal and vertical axis stand for the evolving time and the
disruption rates respectively, with the units of $G=M=a=1$. The solid, dashed and dotted lines stand for the integrations with
$\gamma=0.5,1.0,1.5$ respectively. \label{fig:rtgamma} }
\end{center}
\end{figure}

Figure~\ref{fig:rtgamma} shows the disruption rates with different $\gamma$ values for both the stationary and the recoiling cases. The
left panel is for stationary integrations and the right panel is for the recoiling SMBH with time duration $t=60-100$. Here we can
not compare the same phase for different $\gamma$ models because of their different evolving times. The results in the
right panel are phase II for model $06$ and phase III for model $09$ and $12$. It can be seen that, for both panels, there are
several times differences between every two adjacent $\gamma$ values.

In order to compare with the work in the literature with higher $V_{\rm ej}$, a simulation with similar initial
parameters comparing with model $06$ has been done in model $10$, where $\gamma=1$ and $V_{\rm ej}=1.1V_{\rm esc}\simeq
1.56$. Our result shows that there are just five particles tidally disrupted at the beginning of recoil, and
following with zero event recorded for the rest of time. Because of the low particle resolution, we can only give a
lower limit to the tidal disruption rate. For a galaxy with mass $M = 4\times 10^{10}\msun$, $\Mbh=4\times 10^{7}\msun$
and $r_{1/2} = 1\kpc$, the kick velocity should be $V_{\rm ej}\sim 1000 \kms$. Thus our lower limit for tidal
disruption rate is $\sim 6\times 10^{-8}\accrate$, which is lower than $\sim 10^{-6}\accrate$ in Figure $2$ of
\citet{kom08b} with similar conditions. For the bound stars, we find that there are only $\sim10^{-3}\Mbh$ stars still
bound to SMBH after the recoil, which is roughly as the same order of magnitude as the result in \citet{mer09}. The
small amount of bound stars, low stellar density outside and poor particle resolution result in the rare disruption
events in computation. To solve this problem, a series of simulations with larger particle numbers are needed.

\section{DISCUSSIONS AND CONCLUSIONS}
\label{dissum}

Supermassive binary black holes in galactic nuclei may under certain conditions get very close and coalesce due to strong
emission of gravitational waves. Because of the anisotropy of GW emission, the coalescence remnant SMBH experiences a strong recoil velocity depending on spins of the original pair of black holes. In this paper we investigate the
interaction and co-evolution of the recoiling SMBHs with their galactic stellar environments, using very large direct $N$-body
simulations with a simplified tidal disruption scheme. It is the first numerical investigation for the evolution of the tidal
disruption by an oscillating SMBHs with unbound stars included. Moreover, we discover a polarization cloud of stars surrounding the
oscillating SMBHs, which is not mentioned in the literature.

As argured by \citet{gua08}, the dynamical evolution of the recoiling SMBH can be divided into three phases. During the
phase I, the trajectory of the recoiling SMBH can be well predicted by Chandrasekhar's dynamical friction theory
\citep{cha43}. The phase II begins when the SMBH's oscillation decayed to the size of galactic core, and the motion of the
SMBH damps slowly. At phase III, the orbital oscillation of SMBH slowly decay to achieve a thermal equilibrium with
surrounding stars. In our simulation, we obtain similar phase I and II as \citet{gua08} have got. Besides, our results
show that the mass fraction of bound stars relative to SMBH is roughly $0.7\%$ at the beginning. This value is
consistent with \citet{gua08} and \citet{mer09}.

In addition to following the dynamical evolution of the recoiling SMBH, we estimate the stellar tidal disruption rates by the recoiling
SMBHs using direct $N$-body simulation together with our simple tidal disruption scheme. Our results show that the stellar tidal
disruptions do not significantly change the dynamical evolution of the recoiling SMBH. The stellar disruption rate by the recoiling SMBH in phase II is $\sim10^{-5} \accrate$, which is about the same order of magnitude of the stellar disruption rate by the stationary
SMBH at a spherical galactic center. However, in phase I, the stellar tidal disruption rate is about an order of magnitude
lower.

In our simulation, a spherical stellar distribution is assumed. This simplified model is not accurate to represent a
real galaxy. For most of real bulges/ellipticals, their stellar distributions are non-spherical. As argued by
\citet{vic07}, the orbital decay time scales of the recoiling SMBHs are prolonged in triaxial galaxies comparing with
spherical galaxies. That is because most of the recoiling SMBHs in triaxial galaxies can not return directly to their
galactic centers, where the influences of dynamical friction are the strongest. For the same reason, a non-spherical
dark matter halo can also significantly increase the decay time of a recoiling SMBH \citep{gue09}. Since most of the
tidal disruption events happened during the SMBHs passing through the galactic centers, the triaxiality may
significantly reduce the tidal disruption rate from unbound stars. To investigate the impacts of triaxiality, a good
solution is to have series of simulations with triaxial stellar distributions. However, it is very complicated to
generate a stationary triaxial distribution for $N$-body simulation, and the prolonged orbital decay time needs more
calculation resources. Here we just follow previous works to investigate the tidal disruptions from the recoiling SMBHs in
spherical stellar distribution. Further works with triaxial stellar distribution will be included in our next paper.

Actually, the triaxiality can not change our results significantly because the influence of triaxiality is moderate in
galaxy models with shallow central density profiles \citep{vic07}. Thus our calculations with $\gamma = 0.5$ will not
significantly departure from the triaxial model. Besides, in most of our simulations, the maximal displacement of
the recoiling SMBH relative to the density center is no more than $1\kpc$. That keeps the trajectory of the SMBH inside
the bulge, where the impact of dark matter halo are very weak comparing with baryonic matter. For this reason, we
neglect the influence of the triaxial dark matter halo which is mentioned by \citet{gue09}.

Our simulation for escaping SMBH does not have enough particle resolution to estimate disruption rates. That needs much
larger $N$-body particle number in the future simulations. Here our integrations are focus on relatively low recoiling
velocity. For the assumption that $M = 4\times 10^{10} \msun$ and $r_{1/2} = 1 \kpc$, recoil velocity in our fiducial model
is $\sim 600~\kms$. This velocity produces the oscillating SMBH an amplitude less than the effective radius of
elliptical/bulge. Our results show that most of tidal disruption events occur when the recoiling SMBH passes through
density center, but not exact the original center because of the orbital precession. The tidal disruption rates around
each apocenter are much lower. An estimate for average tidal disruption rate at about apocenters in phase I over
several periods for model $06$ has been done. We calculate the average disruption rate when SMBH have large distant
relative to density center. The disruption rate at large distance is several times or even an order of magnitude lower
than the average value over whole phase I, which implies that the rate for apocenters is one or two order of magnitude
lower than that by stationary SMBHs.

\citet{kom08b} argued that off-nuclear X-ray tidal flares could be one of the observational signatures for an escaping
SMBH. Unfortunately, our simulation of model $10$ with $1\rm M$ particles and the escapable recoil velocity does not have sufficient resolution to estimate the disruption rate. We can just give a lower limit $\sim 6\times 10^{-8}\accrate$, which does not conflict to their work with similar configurations. Besides, our model $09$ with $\gamma=1.0$ and recoil velocity $\sim 600\kms$ corresponds to a tidal disruption rate $\sim 10^{-5}\accrate$. This value is similar to the result of Model 1 with $V_{\rm ej}=500\kms$ in \citet{kom08b}. Based on our results, an oscillating SMBH with enough large recoil velocity may also contribute to off-nuclear X-ray flares around its
apocenters with the low disruption rates.

An HCSS around the recoiling SMBH in our integrations has similar mass as predicted by \citet{mer09}. However, we can
not give more information about the star clusters limited by particle resolution. In addition, we find a larger cloud
of stars around the recoiling SMBHs, which are not permanently bound to the black hole but keep a kinematic correlation
to it for more than an orbital time. It can be described as polarization cloud reacting to the gravitational
disturbance by the recoiling SMBH, in other words, the echo of dynamical friction. The size and mass of the
polarization clouds are comparable to ultracompact dwarf galaxies, except significantly higher velocity dispersion. The
existence of such kind of system may bring difficulties for distinguishing the HCSS. More detailed properties of
the polarization cloud will be studied in our following work.

Our simulations for different particle numbers show that both $250\rm K$ and $1\rm M$ integrations are sufficient in
following the dynamical evolution of the recoiling SMBH, but the former do not have enough resolution for investigating the
tidal disruption and co-evolution of SMBH with surrounding stars. Different initial masses of the recoiling SMBHs can
change the results moderately. We find that a higher galactic density slope can significantly speed up the dynamical
evolution of the system, and the tidal disruption rate is also boosted.

\acknowledgments

Support for this work was provided by the Chinese national $973$ program ($2007CB815405$), the National Natural Science
Foundation of China ($11073002$), the Research Fund for the Doctoral Program of Higher Education (RFDP), and the China
Scholarship Council for financial support ($2009601137$). RS and PB acknowledge support by NAOC CAS through the Silk
Road Project and (RS) through the Chinese Academy of Sciences Visiting Professorship for Senior International
Scientists, Grant Number $2009S1-5$. We thank for partial support by Global Networks and Mobility Program of the
University of Heidelberg (ZUK 49/1 TP14.8 Spurzem). The special supercomputer at the Center of Information and
Computing at National Astronomical Observatories, Chinese Academy of Sciences, funded by Ministry of Finance of
People's Republic of China under the grant $ZDYZ2008-2$, has been used for some of the largest simulations. The titan
cluster has been used, funded by the Volkswagen Foundation under grant No. $I/84 678-680$. Some numerical computation
has also be done on the SGI Altix $330$ system at the Astronomy Department, Peking University. We are grateful to X.-B. Wu, Zhaoyu Li and Heling Yan for useful discussions. Many thanks are due to the referee for valuable comments.

\begin{deluxetable}{cccccc}
    \tablewidth{0pt}
    \tabletypesize{\scriptsize}
    \tablecaption{Parameters of simulation models \label{tab:para}}
    \tablehead{
    \colhead{$Model \, No.$} &
    \colhead{$N$} &
    \colhead{$r_{\rm t}$} &
    \colhead{$\gamma$} &
    \colhead{$\Mbh/M$} &
    \colhead{$V_{\rm ej}/V_{\rm esc}$} \\
    \colhead{(1)} &
    \colhead{(2)} &
    \colhead{(3)} &
    \colhead{(4)} &
    \colhead{(5)} &
    \colhead{(6)}
    }

    \startdata

01  & $50\rm K$ & $5\times 10^{-4}$ & 0.5 & 0.001  &  0.7  \\
02  & $250\rm K$ & $5\times 10^{-4}$ & 0.5 & 0.001  &  0.7  \\
03  & $250\rm K$ & $1\times 10^{-3}$ & 0.5 & 0.001  &  0.7  \\
04  & $250\rm K$ & $5\times 10^{-3}$ & 0.5 & 0.001  &  0.7  \\
05  & $1\rm M$ & $5\times 10^{-4}$ & 0.5 & 0.001   &  0.0  \\
06  & $1\rm M$ & $5\times 10^{-4}$ & 0.5 & 0.001   &  0.7  \\
07  & $1\rm M$ & $5\times 10^{-4}$ & 0.5 & 0.002   &  0.7  \\
08  & $1\rm M$ & $5\times 10^{-4}$ & 1.0 & 0.001   &  0.0  \\
09  & $1\rm M$ & $5\times 10^{-4}$ & 1.0 & 0.001   &  0.7  \\
10  & $1\rm M$ & $5\times 10^{-4}$ & 1.0 & 0.001   &  1.1  \\
11  & $1\rm M$ & $5\times 10^{-4}$ & 1.5 & 0.001   &  0.0  \\
12  & $1\rm M$ & $5\times 10^{-4}$ & 1.5 & 0.001   &  0.7  \\
13  & $1\rm M$ & $5\times 10^{-2}$ & 0.5 & 0.001   &  0.0  \\
14  & $1\rm M$ & $5\times 10^{-2}$ & 0.5 & 0.001   &  0.7  \\
15  & $1\rm M$ & $5\times 10^{-3}$ & 0.5 & 0.001   &  0.0  \\
16  & $1\rm M$ & $5\times 10^{-3}$ & 0.5 & 0.001   &  0.7  \\
17  & $1\rm M$ & $1\times 10^{-3}$ & 0.5 & 0.001   &  0.0  \\
18  & $1\rm M$ & $1\times 10^{-3}$ & 0.5 & 0.001   &  0.7  \\
19  & $1\rm M$ & $5\times 10^{-5}$ & 0.5 & 0.001   &  0.0  \\
20  & $1\rm M$ & $5\times 10^{-5}$ & 0.5 & 0.001   &  0.7  \\

    \enddata

\tablecomments{ Col.(1): Model sequence number. Col.(2): Particle numbers adopted in calculations. Col.(3): Tidal disruption
radius $r_{\rm t}$ in the units $G=M=a=1$. Col.(4): Density slope $\gamma$. Col.(5): Initial BH mass in the unit of total mass.
Col.(6): Initial recoil velocity in the unit of escape velocity. All the simulations set $\epsilon=10^{-5}$ and $\eta=0.01$.}
\end{deluxetable}

\end{document}